\documentclass{ar2e}
\usepackage{ulem}  
\usepackage{color}
\usepackage{graphicx}
\begin{document}

\renewcommand{\baselinestretch}{1.3}

\newcommand{\Red}{\color [rgb]{0.8,0,0}}
\def\SSG#1{{\Red [SSG: #1]}}
\newcommand{\Green}{\color [rgb]{0,0.7,0}}
\def\AK#1{{\Green [AK: #1]}}

\def\mop#1{\mathop{\rm #1}\nolimits}
\def\Vol{\mop{Vol}}
\def\Re{\mop{Re}}
\def\Im{\mop{Im}}
\def\overleftrightarrow#1{\vbox{\ialign{##\crcr
     $\leftrightarrow$\crcr\noalign{\kern-0pt\nointerlineskip}
     $\hfil\displaystyle{#1}\hfil$\crcr}}}
\def\lsim{\mathrel{\mathstrut\smash{\ooalign{\raise2.5pt\hbox{$<$}\cr\lower2.5pt\hbox{$\sim$}}}}}
\def\gsim{\mathrel{\mathstrut\smash{\ooalign{\raise2.5pt\hbox{$>$}\cr\lower2.5pt\hbox{$\sim$}}}}}

\newcommand{\diag}{{\rm diag\,}}

\ARinfo{\hskip4.4in PUPT-2276}

\title{From gauge-string duality to strong interactions: a Pedestrian's Guide}

\markboth{Gubser and Karch}{Pedestrian's Guide}

\author{Steven S. Gubser
\affiliation{Joseph Henry Laboratories, Princeton University, Princeton, NJ  08544}
Andreas Karch
\affiliation{Department of Physics, University of Washington, Seattle, WA  98195-1560}}

\begin{keywords}
Quantum chromodynamics, string theory, gauge-string duality
\end{keywords}

\begin{abstract}
We survey recent progress in understanding the relation of string theory to quantum chromodynamics, focusing on holographic models of gauge theories similar to QCD and applications to heavy-ion collisions.
\end{abstract}

\maketitle

\setlength{\baselineskip}{18pt}

\section{Introduction}
\label{INTRODUCTION}

Quantum chromodynamics (QCD) \cite{Gross:1973id,Politzer:1973fx} is the accepted theory of strong nuclear forces.  But QCD is hard to solve.  Perturbation theory gives access to physics at high scales, well above the confinement transition.  Lattice gauge theory is best suited for the calculation of static quantities such as the vacuum structure, the spectrum, and thermodynamics.  Various heuristic models and approximation schemes, ranging from the bag model to fluid dynamics, supply partial information about the dynamics of QCD in the strongly coupled regime.

Over the past ten years or so, string theory has supplied a new way of thinking about strongly coupled gauge theories: the gauge-string duality \cite{Maldacena:1997re,Gubser:1998bc,Witten:1998qj}.  One of the simplest cases is maximally supersymmetric Yang-Mills theory, which is understood to be dual to string theory in a specific curved ten-dimensional geometry, $AdS_5 \times S^5$.  The gauge-string duality can also handle confining gauge theories which are at least qualitatively similar to QCD.

As compared to the more standard tools for understanding QCD, the gauge-string duality offers two main advantages.  First, it offers direct access to the strong coupling region.  The simplest examples, like maximally supersymmetric Yang-Mills theory, are perfectly conformal, with identically zero beta function.  But they are nevertheless strongly interacting, and confining behavior is in some sense just around the corner.  Second, it is straightforward to introduce finite temperature without passing to Euclidean signature: one instead introduces a black hole horizon.  This makes it relatively straightforward to extract transport coefficients, such as viscosity and diffusion constants, for analogs of the quark-gluon plasma. All these real time phenomena are challenging to calculate on the lattice; the complex weighting factor $e^{i S}$ where $S$ is the action in the path integral makes the standard importance sampling used on the lattice unreliable.

The gauge-string duality also suffers from several deficiencies as a phenomenological tool.  The worst one is that we do not have a dual for QCD itself.  We are therefore forced to make comparisons in a more or less ad hoc fashion between real-world QCD and the theories we can deal directly with via the gauge-string duality.  Another serious problem is that the strong coupling limit is taken quite strictly in many or most gauge-string calculations: only the leading behavior in the limit of infinite coupling is readily accessible.  Significant inroads have been made on computing subleading corrections, but both technical and interpretational challenges are substantial.  Likewise, it is in general difficult to proceed past the leading order in large $N$ expansion, where $N$ is the number of colors.

The role of the gauge-string duality in understanding strong interactions is still evolving.  The duality has made an indelible mark on our understanding of QCD, and it offers uniquely geometrical perspectives on confinement and the quark-gluon plasma.  But it hasn't solved QCD.

The rest of this review is organized as follows.  In section~\ref{AdSCFT} we introduce the basic ideas of the gauge-string duality.  In section~\ref{flavor} we summarize how flavor is included by introducing branes into negatively curved bulk geometries in the probe approximation.  In section~\ref{CONFINING} we outline gauge-string constuctions that capture aspects of confinement.
In section~\ref{VISCOSITY} we present gauge-string calculations of shear and bulk viscosity of strongly coupled, deconfined plasmas.  In section~\ref{EXPANDING} we review work on expanding plasmas in the dual supergravity description.  In section~\ref{PROBES} we summarize several approaches to the physics of hard probes of a thermal medium.  We conclude briefly in section~\ref{CONCLUDE}.

The size of the relevant literature and the brevity of this review have prevented us from discussing many of the topics with the thoroughness they deserve.  Also, we have altogether omitted some important developments, notably the work on anomalous dimensions of long operators (see for example \cite{Beisert:2003yb}); string descriptions of gluon scattering \cite{Alday:2007hr};  bottom-up approaches to holographic models of confinement (see for example \cite{Erlich:2005qh,DaRold:2005zs,Karch:2006pv}); and light-front holography, reviewed for example in \cite{Brodsky:2008kp}.

\section{The basics of the gauge-string duality}
\label{AdSCFT}

The story of the gauge-string duality turns on understanding D3-branes.  There are other types of branes in string theory, but D-branes are the best understood, thanks to the crucial contribution of Polchinski \cite{Polchinski:1995mt}. By definition, D-branes are locations in ten-dimensional spacetime where strings can end.  A D3-brane is a D-brane extending in three spatial dimensions as well as in the time direction.  The low-energy excitations of a single D3-brane are described by ${\cal N}=4$ supersymmetric $U(1)$ gauge theory.  If $N$ D3-branes are placed on top of one another, one finds instead ${\cal N}=4$ supersymmetric $U(N)$ Yang-Mills gauge theory \cite{Witten:1995im} (hereafter ${\cal N}=4$ SYM).  This theory splits into a $U(1)$ part, which is free, and relates to the center of mass motions of the stack of D3-branes; and an $SU(N)$ part, which is interacting, and relates to the relative motions of the branes.  The gluons of these gauge theories can be represented as strings running from one brane in the stack to another.  If $N=3$ and we label the branes as $R$, $G$, and $B$, then a string running from $R$ to $B$ has color quantum numbers $\bar{R}B$, as appropriate for a gluon of $SU(3)$ gauge theory.  The ${\cal N}=4$ supersymmetry of D3-branes dictates that this $\bar{R}B$ string can act not only as a gluon, but also as spin-$1/2$ fermion or as a scalar; the strings describing these superpartners of the gluons differ from the ones describing the gluons themselves by fermionic excitations on the string. Regardless of spin, all particles in ${\cal N}=4$ SYM are charged in the adjoint of the gauge group.  The precise dynamics of the gauge theory is almost entirely dictated, at the level of a renormalizable lagrangian, by the number of branes $N$ and the supersymmetry.  The only freedoms one has are to adjust the gauge coupling and a $\theta$ parameter.  The $\theta$ parameter breaks CP, and we set it to zero from here on.  This amounts in the D3-brane construction to requiring that a particular scalar field (more precisely, a pseudoscalar) should vanish.

D3-branes have a definite mass per unit volume, and they have a definite charge under a 5-form field strength.  They therefore deform spacetime into a solution of the ten-dimensional Einstein equations coupled to the 5-form.  Close to the D3-branes, this solution takes the form of a direct product, $AdS_5 \times S^5$, where $AdS$ stands for anti-de Sitter space.  The gauge-string duality \cite{Maldacena:1997re,Gubser:1998bc,Witten:1998qj} is an equivalence between string theory on this geometry and ${\cal N}=4$ SYM in four dimensions.  Green's functions of gauge-invariant operators, Wilson loops \cite{Rey:1998ik,Maldacena:1998im}, and thermal states \cite{Gubser:1996de,Witten:1998zw} are just some of the gauge theory constructions which have well-studied translations into the language of string theory in anti-de Sitter space.

To get started with AdS/CFT, the first step is to understand the metric of extremal D3-branes.  Extremal in this context means that the D3-branes are at zero temperature, so they are described by a vacuum state of ${\cal N}=4$ super-Yang-Mills theory.  The metric is
 \begin{equation}\label{DthreeMetric}
  ds_{10}^2 = H^{-1/2} (-dt^2 + d\vec{x}^2) + H^{1/2}
    (dr^2 + r^2 d\Omega_5^2) \,,
 \end{equation}
where
 \begin{equation}\label{WhatIsH}
  H = 1 + {L^4 \over r^4} \,,
 \end{equation}
and $L$ is related to the number of D3-branes:
 \begin{equation}\label{WhatIsL}
  L^4 = {\kappa \over 2\pi^{5/2}} N = g_{YM}^2 N \alpha'^2 \,.
 \end{equation}
Here $\kappa = \sqrt{8\pi G_{10}}$ is the gravitational coupling in ten-dimensional supergravity, $g_{YM} = \sqrt{4\pi g_{\rm string}}$ is the gauge coupling of ${\cal N}=4$ super-Yang-Mills, $N$ as before is the number of D3-branes, and $\alpha'$ is the Regge slope parameter of fundamental strings, inversely proportional to their tension.  Often we will use the notation $\lambda = g_{YM}^2 N$ for the 't~Hooft coupling.

The region where $r \gg L$ is asymptotically flat ten-dimensional spacetime, ${\bf R}^{9,1}$.  The D3-branes can be thought of as a defect in that spacetime, extended in the $t$ and $\vec{x}$ directions, and having a characteristic size $L$ in the other six directions.

The crucial technical step of \cite{Maldacena:1997re} is to focus on the region $r \ll L$ by dropping the $1$ from the function $H$.  The resulting geometry is a direct product of $AdS_5$ and $S^5$:
 \begin{equation}\label{ProductMetric}
  ds_{10}^2 = {r^2 \over L^2} (-dt^2 + d\vec{x}^2) +
    {L^2 \over r^2} dr^2 + L^2 d\Omega_5^2 \,.
 \end{equation}
Physically, dropping the $1$ from $H$ means we're zooming in close to the D3-branes: so close that we lose track of the asymptotically flat region.  In short, (\ref{ProductMetric}) is a way of replacing the branes by a curved geometry, and the essential claim of AdS/CFT is that the gauge theory dynamics built from strings on the branes is equivalently captured by the geometry (\ref{ProductMetric}).

A variety of different choices of radial variable are in common use.  Among the most common are the following two:
 \begin{equation}\label{OtherVars}
  z = {L^2 \over r} \qquad\qquad u = {r \over L^2} \,.
 \end{equation}
Note that $z$ has dimensions of length, while $u$ has dimensions of energy.  Using the $z$ coordinate makes it clear that $AdS_5$ is conformal to flat space:
 \begin{equation}
  ds_5^2 = {L^2 \over z^2} (-dt^2 + d\vec{x}^2 + dz^2) \,.
 \end{equation}
More precisely, $AdS_5$ is conformal to the $z>0$ half of five-dimensional flat space, ${\bf R}^{4,1}$.  It has a boundary at $z=0$, and that boundary is Minkowski space, ${\bf R}^{3,1}$, parametrized by $t$ and $\vec{x}$.  A common way of speaking is that the gauge theory lives on this boundary, while string theory lives in the bulk of $AdS_5$.  This point of view is useful, but it is perhaps more accurate to say that string theory on the whole of $AdS_5 \times S^5$ is equivalent to ${\cal N}=4$ super-Yang-Mills on ${\bf R}^{3,1}$.

To get a feel for the geometry of $AdS_5$, consider the following two thought experiments:
 \begin{enumerate}
  \item An observer at fixed depth $z_0$ shoots a light ray straight upward toward the boundary.  What happens?  Well, the light ray travels upward along a trajectory $z(t) = z_0 - t$, so it reaches the boundary in a time $t=z_0$.  At least some of the light presumably is reflected back down to the observer, who receives it at a time $t=2z_0$.  This is all exactly as it would be if the observer lived in the $z>0$ half of ${\bf R}^{4,1}$.
  \item The observer extends a rod upward toward the boundary of $AdS_5$.  His rod is very long.  He notices however that its tip never hits the boundary.  What's going on?  Well, the proper distance between the observer and the boundary is infinite, because it is given by the integral
 \begin{equation}
  \ell = \int_0^{z_0} dz \, {L \over z} \,.
 \end{equation}
The divergence comes from the region near $z=0$.  Evidently, this is very unlike the situation in the $z>0$ half of ${\bf R}^{4,1}$.
 \end{enumerate}
In summary, $AdS_5$ gets very, very big near the boundary, but light can nevertheless get to the boundary in finite time.

A crucial ingredient in the relation of $AdS_5 \times S^5$ to ${\cal N}=4$ SYM is the action of dilations.  First, on the field theory side, the beta function of ${\cal N}=4$ SYM vanishes identically.  So the renormalization group says that dilations act trivially on the theory: $T^m{}_m = 0$ as an operator identity.  If we turn to the gravity side, a dilation should be just a coordinate transformation, $x^m \to K x^m$, where $x^m = (t,x,y,z)$ and $K$ is a constant.  By itself, this coordinate transformation does not preserve the form of the metric (\ref{ProductMetric}).  But if we also send $r \to r/K$, or equivalently $u \to u/K$ or $z \to K z$, then the metric is preserved.  Thus we have associated dilation symmetry of ${\cal N}=4$ SYM with an isometry of the $AdS_5$ metric. Taking $K > 1$ means making things bigger in the $x^m$ directions.  At the same time, $u \to u/K$ means that the location in the radial direction gets smaller---``deeper'' in $AdS_5$, and further from the boundary.

The upshot of this discussion is that the region of $AdS_5$ where $u$ is large corresponds to ultraviolet (UV) physics, whereas the region where $u$ is small corresponds to infrared (IR) physics.  A slight generalization of the $AdS$ part of the metric (\ref{ProductMetric}) helps clarify this:
 \begin{equation}
  ds^2 = L^2 \left[ u^2 (-h dt^2 + dx_1^2 + dx_2^2 + dx_3^2) +
    {du^2 \over h} \right] + L^2 d\Omega_5^2 \,,  \label{AdSSch}
 \end{equation}
where $h = 1 - u_h^4/u^4$.  This geometry satisfies the Einstein equations with the same five-form field strength that supported the original $AdS_5 \times S^5$ metric.  The solution (\ref{AdSSch}) describes the near-horizon geometry of near-extremal D3-branes.  Near-extremal means that a little bit of mass has been added to the D3-branes, in the form of thermal energy.  The temperature manifests itself as the Hawking temperature of the horizon at $u=u_h$, which is $T = u_h/\pi$.  Thus we can regard $u$ or $u/\pi$ as a typical energy scale pertaining to processes taking place in $AdS_5$ at a depth $u$.

To understand the calculations of Green's functions in the gauge-string duality, it helps to return for a moment to the idea of a D3-brane as a defect with transverse size $L$ in asymptotically flat ten-dimensional spacetime.  In order to explore its physical properties, the first thing we might try is throwing things at it and seeing how they bounce off or are absorbed.  If not for the fact that the D3-brane extends infinitely far in the three directions parametrized by $\vec{x}$, we could assign cross-sections to processes of this type.  Because of its infinite extent, the right quantity to consider is the cross-section per unit volume in the $\vec{x}$ directions.  There is a large literature devoted to exploring D-branes in this way,  including \cite{Klebanov:1995ni,Das:1996wn,Callan:1996dv,Das:1996we,Klebanov:1997kc,Gubser:1997se}, and the gauge-string correspondence is in part an outgrowth of it.  The typical process by which a D3-brane absorbs a closed string is to split it into two or more open strings that propagate along the worldvolume of the branes.  If the closed string is a graviton, such a process is mediated by the coupling $h_{mn} T^{mn}$, where $T^{mn}$ is the stress tensor of ${\cal N}=4$ SYM.  Then the open strings are gluons or superpartners of gluons.  From the field theory point of view, the calculation of the response of the brane to the perturbation by an external metric amounts to the calculation of correlation functions of the stress tensor.

To compute the absorption cross-section of a graviton by D3-branes, one is interested in an inclusive process where the final state is arbitrary.  The optical theorem guarantees a relation between the cross-section of the inclusive process and the two-point function of the vertex operator $T^{mn}$.  The two-point function characterizes the linear response of the D3-branes to a metric perturbation.  Higher point functions would characterize this response past the linearized approximation.  The calculation of Green's functions \cite{Gubser:1998bc,Witten:1998qj} can be understood, in part, as a streamlined treatment of absorption processes.  Here is a heuristic summary how gauge theory Green's functions are calculated using $AdS_5 \times S^5$.  A graviton perturbation $h_{mn}$ creates an environment that the D3-branes respond to.  But in the gravitational picture, the near-horizon geometry (i.e. $AdS_5 \times S^5$) effectively replaces the D3-branes.  So the graviton perturbation creates an environment that $AdS_5 \times S^5$ responds to.  That environment should be thought of as imposed at some very large value of $u$, because that is where $AdS_5 \times S^5$ matches onto asymptotically flat space.  So the prescription should be to impose some metric deformation in the limit of large $u$ and then carry out the path integral of string theory in $AdS_5 \times S^5$ subject to boundary conditions corresponding to the deformation.  The result of the path integration then has to be the generating functional of Green's functions for color-singlet operators in the field theory.  If the metric is the only field one deforms at large $u$, then one has access only to $n$-point functions of $T^{mn}$.  But there is a large list of additional operators for which dual fields in $AdS_5 \times S^5$ are well understood.

A big stumbling block is that the path integral of string theory in $AdS_5 \times S^5$ is poorly understood.  The best-studied approximation to it is the saddle point expression $e^{iS}$, where $S$ is the gravitational action, classically extremized subject to appropriate boundary conditions.  This approximation is good when $AdS_5$ has small curvature compared both to the Planck scale and to the string scale.  The former constraint turns out to mean that the number of D3-branes must be large: $N \gg 1$.  The latter means that the 't Hooft coupling must be large: $g_{YM}^2 N \gg 1$.

\section{The addition of fundamental matter}
\label{flavor}

One ingredient that our discussion so far has been missing is fundamental matter.  The quarks of QCD are fermions transforming in the fundamental representation of the $SU(3)$ gauge group.  They interact via exchange of gluons, which are massless spin-$1$ bosons transforming in the adjoint representation of $SU(3)$.  The ${\cal N}=4$ theory describes supersymmetric glue, in which all particles transform in the adjoint.  So we are still missing the quarks. In the limit of a large number of colors that underlies the gauge/string correspondence, the difference in representation is crucial. Topologically, the fundamental representation quarks introduce boundaries into the double-line Feynman diagrams. Each fundamental matter loop suppresses the diagram by a power of $N$ in the large $N$ counting. For example, the free energy of a free gas of gluons and quarks has an order $N^2$ contribution from the gluons and an order $N_f N$ contribution from the quarks, where $N_f$ is the number of different flavors of quarks. The naive large $N$ counting will work as long as $N_f \ll N$. For this very basic $N$ counting the spin of the quarks is irrelevant; for that reason it has become common in the gauge/string literature to refer to all fields in the adjoint representation collectively as ``the glue'' and all fundamental representation fields as ``the quarks'' or ``the flavors'', irrespective of spin. One important consequence of this large $N$ counting is that dynamical fermion loops are suppressed at large $N$. Inclusion of the flavor fields allows for new observables, e.g. the chiral condensate, the meson spectrum, flavor current correlators or the order $N$ contribution to the free energy. But all internal lines in the diagrams are only glue lines. The quark dynamics is happening in the background of the strongly coupled glue. One can for example even add additional fundamental matter to the conformal ${\cal N}=4$ SYM. This will make the $\beta$ function positive and asymptotic freedom is lost. The coefficient of the $\beta$ function however is of order $1/N$ and so this conformal symmetry breaking and the appearance of the Landau pole can be ignored at large $N$.

These statements about Feynman graphs have of course also a direct analog on the string theory side. Since in the topological counting fundamental matter introduces boundaries, the string theory dual will have to have open strings, too. Including open strings amounts to the addition of D-branes in the bulk, the places where strings can end.  D-branes in string theory have a tension of order one over the coupling, that is of order $N$ in the limit of large $N$ at finite $g_s N$; the stress tensor $T^{\rm D-brane}_{\mu \nu}$ hence also is of order $N$. Newton's constant $G_N$ on the other hand is of order $g_s^2 \sim N^{-2}$. In the presence of D-branes Einstein's equations need to be re-solved as the additional D-branes curve space. However the right hand side is of order $G_N \, T^{\rm D-brane}_{\mu \nu} \sim 1/N$ and hence vanishes in the large $N$ limit. The classical dynamics of these ``flavor D-branes'' will demand that they minimize their action in the presence of a given background geometry. This essentially boils down to a minimal area problem, as the action of a D-brane is proportional to its worldvolume.  (In some situations it is important to know that there is a gauge field on the brane, unrelated to color.  Its inclusion leads to the Dirac-Born-Infeld action or DBI action.  There are also fermionic fields on the brane.)  As in the field theory the dynamics of the glue, which is encoded in the background geometry, is unaltered by the presence of the flavor branes at leading order in $N$. The first example of such a flavored version of AdS/CFT was a conformal example with massless quarks but with the backreaction fully included \cite{Aharony:1998xz}. The realization \cite{Karch:2000gx,Karch:2002sh} that flavors, massless and massive, can be studied in the ``probe limit,'' that is in the limit where the backreaction of the brane can be neglected, opened the door to a systematic study of the flavor sector in AdS/CFT.

In order to make these ideas more explicit, let's consider in more detail a D7-brane probe in $AdS_5 \times S^5$, following \cite{Karch:2002sh}.  It is useful to begin with what seems like a detour: consider the D7-brane and the D3-branes both in the probe approximation in ten-dimensional flat space.  Let's assume that the D7-brane is extended along the same spatial directions as the D3-branes, and that the distance of closest approach between the D7 and the D3's is $r_{\rm D7}$.  It is a textbook exercise to show that eight supersymmetries are preserved by such a configuration---provided we leave all the branes at zero temperature.  The dynamics of low-energy excitations, including strings running between the D3-branes and the D7-brane, is completely determined by the supersymmetry.  The D3-D7 strings, in particular, are described in terms of massive scalars and fermions, fundamentally charged under the $SU(N)$ gauge group on the D3-branes.  Their mass is $m = r_{\rm D7}/2\pi\alpha'$, simply because this is the tension of a string with length $r_{\rm D7}$.

With fully a quarter of supersymmetry preserved, one expects that the behavior of the D7-branes will scarcely be altered if one replaces the probe D3-branes by the warped metric (\ref{DthreeMetric}).  The $dr^2 + r^2 d\Omega_5^2$ part of this metric is the six-dimensional flat space transverse to the D3-branes.  The D3-branes are located at the origin of this transverse space.  The simplest embedding of the D7-branes is to have them run along a four-dimensional plane in the transverse space whose closest approach to the origin is at a distance $r_{\rm D7}$.  If we express the metric of the unit $S^5$ as
 \begin{equation}
  d\Omega_5^2 = d\theta^2 + \cos^2 \theta \, d\psi^2 +
    \sin^2 \theta \, d\Omega_3^2 \,,
 \end{equation}
where $d\Omega_3^2$ is the metric of a unit $S^3$, then the four-dimensional plane can be specified by the equations $\psi=0$, $r\cos\theta = r_{\rm D7}$. 
It can be checked that this embedding does extremize the D7-brane worldvolume action, which is proportional to its area.  The mass of a string running from the D3-branes to the D7-brane is still $m = r_{\rm D7}/2\pi\alpha'$, because in the string world-volume action, a factor of $H^{1/4}$ from the string's extension in a transverse direction cancels against a factor of $H^{-1/4}$ from its propagation in the time direction.

To achieve a description of a D7-brane probe in $AdS_5 \times S^5$, one can rely upon the previous discussion, but simply drop the $1$ from the harmonic function $H$.  Geometrically, the D7-brane wraps an $S^3$ in the $S^5$, and it extends across a range of radii in $AdS_5$, $r \geq r_{\rm D7}$.  A string stretched between the D3-branes to the D7-brane is now understood as a string running from the depths of $AdS_5$, where $r \to 0$, up to $r=r_{\rm D7}$.  One may also consider as a limiting case $r_{\rm D7}=0$, which means that the D7-brane passes right through the D3-branes.  Then the strings running from the D3-branes to the D7-brane act as massless fundamentally charged particles.  In this special case, the D7-brane wraps an equatorial $S^3$ in $S^5$.  It might appear that an instability should arise from perturbations where the D7-brane slips away from the equator to an $S^3$ of smaller size.  In fact, this mode corresponds to a tachyon of mass $m^2 = -3/L^2$, but according to the analysis of \cite{Breitenlohner:1982jf}, tachyons do not indicate an instability in $AdS_5$ until $m^2$ falls below $-4/L^2$.  Heuristically, this is because $AdS_5$ acts as a box, and fluctuations must always go to zero at its boundary.  The lowering of potential energy one achieves by exciting a tachyonic mode is offset, provided $m^2 \geq -4/L^2$, by the raising of kinetic energy due to this boundary condition.

The complete meson spectrum in the example of the D3/D7 system was worked out in \cite{Kruczenski:2003be}.  Mesons are the eigenmodes of small gauge field fluctuations of the brane. Gauge field propagation on the brane in response to an excitation by an external global current calculates current correlation functions in the dual field theory. This is in direct analogy with the relation between $n$-point functions of the stress tensor in the field theory and the analysis of the response of the
$AdS_5$ geometry to an external gravitational perturbation we described before. For massive quarks this correlation function is dominated by discrete modes, the mesons.

Other types of flavor can be constructed by starting with D-branes of different dimensions.  Let there be $N$ coincident D$p$-branes, which we refer to as ``color branes'' because the gauge theory on their world-volume is thought of as the one analogous to $SU(3)$ color.  Let there be one (or $N_f \ll N$) D$q$-branes, called ``flavor branes,'' which are treated in the probe limit.  The field theory description of such a setup depends crucially on one parameter: the number of so-called ``ND'' directions.  An ND direction is a direction which is parallel to the world-volume of the flavor branes, but orthogonal to the world-volume of the color branes, or vice versa.  In the D3/D7 example above, there are four ND directions, because the D7-branes extend in four directions that the D3-branes do not.  Systems with four ND directions preserve eight supersymmetries, which essentially fixes the Lagrangian.  Another interesting class of flavor branes has six ND directions. In this case all supersymmetry is broken. The fundamental matter that is added is just a single Weyl fermion. We will discuss one example of this type below, the Sakai-Sugimoto model.

\section{Gravity Duals of Confining Gauge Theories}
\label{CONFINING}
\subsection{The Hard Wall paradigm}
\label{paradigm}

As scale transformations map to an isometry of $AdS_5$, it is clear that in order to describe a confining gauge theory we need to move beyond $AdS_5$ in the dual gravity.  More particularly, energy scale in the field theory is encoded in the radial direction of the dual gravitational theory, so there will be a spatial radial coordinate $z_*$ that maps to the strong coupling scale $\Lambda_{\rm QCD}$. In pure $AdS_5$, plane-wave normalizable fluctuations around the background can come with any timelike four-momentum $p^2$.
These modes should be thought of as the particle spectrum of the dual field theory.  This continuous spectrum is completely appropriate for the dual of a conformal field theory without well defined particles. For a confining gauge theory with a discrete excitation spectrum, the dual gravity side should also only have discrete modes for the fluctuations with four momentum $p^2 = m_n^2$ with discrete masses $m_n$. This can easily be achieved if the extra dimension truncates around $z_*$ instead of extending all the way to $z=\infty$. Only a discrete set of modes will satisfy
both the boundary condition at $z=0$ as well as at $z_*$. The details of the confinement mechanism are encoded in the geometry around $z_*$. Many generic features of confining theories with a gravity dual do not depend on these details and can be analyzed by simply truncating undeformed $AdS_5$ at $z_*$ as first advocated in \cite{Polchinski:2001tt}. This is often referred to as a ``hard wall.'' In the following, we will review several string theory constructions that realize this hard wall paradigm. In most of the known examples the geometry is actually quite distinct from $AdS_5$ even in the UV. Only one of the examples we will discuss becomes $AdS_5$ in the UV.

Despite the fact that these are confining gauge theories, the confining theories with gravity dual are still very different from QCD. In the original conformal example of AdS/CFT we need to work in the limit of large 't Hooft coupling in order to be allowed to approximate the classical bulk string theory by classical supergravity. One way to understand this is that large 't Hooft coupling ensures a large hierarchy between the mass of the excited modes of the string (governed by the string length) and the light modes of the string (governed by the curvature radius). The confining theories also have to have this large hierarchy of scales in order for classical gravity calculations to be applicable.  Again the string scale on the gravity side (dual to the flux tube tension in the field theory) is much larger than the typical energies of the gravitational modes (which map to the mass gap of the field theory). A typical confining gauge theory will have a flux tube tension $\sigma$ set by the same scale $\Lambda_{\rm QCD}$ as the mass gap.  In short, a confining theory with a gravity dual needs a large dimensionless number $\lambda \sim \sigma^2/\Lambda^4_{\rm QCD}$.

Confining theories with gravity dual share many common qualitative features which can be understood in terms of the hard wall paradigm. Most importantly, Wilson loops will exhibit area law behavior.  To calculate the Wilson loop one needs to identify the minimal area worldsheet connecting the two test charges at the AdS boundary \cite{Maldacena:1998im,Rey:1998ik}.  As for all other correlation functions, this worldsheet is the response of the gravity system to a perturbation imposed from the outside: in this case a quark and anti-quark pair that remain at fixed separation for a specified time. In the conformal case, the worldsheet sinks down deeper and deeper into the bulk as the external quarks are separated along the field theory directions. In a theory with any kind of hard wall, the string can sink no further than $z_*$. For large separation between the test quarks the minimal area is hence dominated by a string lying at the bottom of the geometry at $z_*$.  The latter corresponds to a flux tube with an effective tension given by $\sqrt{g_{tt}(z_*) g_{xx}(z_*)}/2\pi\alpha'$, where $1/2\pi\alpha'$ is the tension of a fundamental string. For ${\cal N}=4$ SYM with a hard wall at $z_* = 1/\Lambda$ this evaluates to $\frac{\sqrt{\lambda}}{2 \pi} \Lambda^2$.  Another property confining gauge theories with gravity dual share with their generic cousins is power laws in elastic hard scattering processes \cite{Polchinski:2001tt} and Regge physics \cite{Brower:2006ea}, that
is scattering at large center of mass energy $s$ at fixed Mandelstam parameter $t$.

One place were the difference between generic confining gauge theory and a confining gauge theory with gravity dual becomes apparent is deep inelastic scattering (DIS) as studied in \cite{Polchinski:2002jw}. DIS requires one to understand the time-ordered two point function of an electromagnetic current in an hadronic state.
At large $q^2$ the matrix element can effectively be calculated using the operator product expansion (OPE). So the main physical question is what operators dominate the current-current OPE.
At weak coupling it is simple single trace operators that dominate the OPE.  In QCD they are bilinear in the quark fields. They create and then annihilate a single parton from the vacuum. At strong coupling these operators generically get an anomalous dimension of order $\lambda^{1/4}$. In the gravitational dual the fields corresponding
to these operators are excited string modes. The dominant operators must be dual to supergravity modes, not string modes. In \cite{Polchinski:2002jw} it was argued that the corresponding field theory operators are double trace operators, that is they are bilinear in hadron fields, and hence create and annihilate a whole hadron, not just a single constituent.
A more intuitive way to describe this result is in terms of the parton content of a hadron. At strong coupling, that is for confining gauge theories with a gravity dual, processes are typically dominated by a very large number of very soft gluons. This stands in stark contrast to a weakly coupled theory such as QCD in the asymptotically free regime where, in a typical process, a few
partons carry order 1 fractions of the total energy and momentum. This basic qualitative insight guides the understanding of many of the strong coupling processes at zero and finite temperature we will describe in this review. Since soft gluons are the main source of difficulty in QCD, gauge theories with supergravity duals that are dominated by soft gluons are a great theoretical laboratory.

\subsection{Mass Deformed ${\cal N}=4$ theories: ${\cal N}=2^*$ and ${\cal N}=1^*$}

Closest among confining backgrounds to the hard wall paradigm described above, with a spacetime that is $AdS_5$ in the UV and ends in
a wall in the IR, is the Polchinski-Strassler (PS) background \cite{Polchinski:2000uf}. On the field theory side the idea is extremely simple: in order to reduce ${\cal N}=4$ SYM to pure glue, add mass terms to the extra scalars and fermions. In the UV the mass terms are inconsequential, so asymptotically the dual spacetime is the standard $AdS_5 \times S^5$ solution
of type IIB string theory. In the IR the heavy fields can be integrated out and one is left with pure glue, which is expected to be a confining theory. Flavor can be added using the technique described in section \ref{flavor}, as was done for the PS background in \cite{Apreda:2006bu}.

How about the hierarchy between flux tube tension and mass gap, which we argued above should be a feature of any confining theory with a gravity dual? When we say that at scales below the mass of the fermions and scalars the theory reduces to pure glue, we have implicitly been assuming that we can take a limit where the mass $m$ is much larger than $\Lambda_{\rm QCD}$.  So what happens if we take this limit $m \gg \Lambda_{\rm QCD}$? At scales below $m$ the theory is pure glue and we know how the coupling evolves with scale. Asymptotic freedom of the pure glue theory guarantees that by the time we let the coupling evolve from $\Lambda_{\rm QCD}$ to $m\gg\Lambda_{\rm QCD}$ the Yang-Mills coupling is small. At energies above $m$ we have the conformal ${\cal N}=4$ theory and the coupling does not evolve any further. So the limit of $m\gg\Lambda_{\rm QCD}$ forces us to consider the ${\cal N}=4$ theory at weak coupling. But this is not where the supergravity analysis is reliable!  Conversely, if one sticks with the strongly coupled ${\cal N}=4$ at energies above $m$ so that supergravity is reliable, one is forced to take $m \sim \Lambda_{\rm QCD}$. The theory one obtains this way does {\it not} reduce to pure glue: it has extra light adjoint fields with masses of order $\Lambda_{\rm QCD}$.  These extra light fields conspire to push the flux tube tension up to a parametrically higher scale than the scale of the mass gap. It is, of course, also possible to only give mass to some of the extra fields leaving one with pure super Yang-Mills in the limit $m \gg \Lambda_{\rm QCD}$, but again with extra light adjoints in the limit that supergravity is valid. The corresponding gauge theories are typically referred to as ${\cal N}=2^*$ and ${\cal N}=1^*$, depending on the amount of unbroken supersymmetry.  The superscript $^*$ is understood to indicate the presence of the extra light adjoint fields.

\subsection{Witten's Black Hole and Sakai-Sugimoto Model}
\label{ss}

One simple way to break both supersymmetry and conformal invariance is to heat a system up to finite temperature.  In Euclidean signature, this corresponds to making time periodic with anti-periodic boundary conditions for the fermions. Long distance questions are dominated by an effective theory, which for both QCD and ${\cal N}=4$ SYM is just pure Yang-Mills theory in 3 real dimensions: the fermions are massive due to the anti-periodic boundary conditions, the scalars inherit this mass via loop corrections and the electric field is heavy due to Debye screening which gives the $A_t$ component a mass of order $g T$. Long distance finite temperature physics is the physics of magnetic fluctuations. The fact that pure glue in 3d confines makes long distance physics at scales $1/(g^2 T)$ non-perturbative even at arbitrary high temperatures in QCD. One can reinterpret these well known facts to achieve the goal of realizing a confining gauge theory with gravitational dual: instead of viewing this system as ${\cal N}=4$ SYM or QCD at finite temperature one can also view it as a 3d gauge theory at zero temperature which was obtained by compactifying ${\cal N}=4$ SYM on a spatial circle of radius $R = \frac{1}{T}$ with anti-periodic boundary conditions for the fermions. At low energies this 3d theory is a theory of pure glue with a dynamical scale $\Lambda \sim g^2_{YM,3d} = g^2_{YM,4d}(R)/R$. There are additional degrees of freedom which come from the higher Fourier modes of the 4d fields on the circle, the so-called Kaluza-Klein (KK) modes. They have a mass of order $1/R$. For a weakly coupled theory (which in the case of QCD means very small $R$) the dynamical scale and the mass scale of the KK-modes are widely separated. In the limit that the 4d coupling and $R$ both go to zero keeping the 3d coupling fixed, the KK-modes completely decouple, and in this limit the theory describes pure 3d glue. It is the opposite limit of strong coupling in 4d in which the theory has a gravity dual. In this limit we have a theory of 3d glue coupled to the tower of KK-modes whose mass is of the same order as the dynamical scale. Again, these additional modes conspire to push the flux tube tension up to a much higher scale than the mass gap.

As we already discussed, in the gravity dual finite temperature corresponds to the introduction of a black hole; the metric is given by the Wick rotation of Eq. (\ref{AdSSch}). The Euclidean black hole geometry relevant for equilibrium physics has the geometry of a cigar: Euclidean time is a circle, and it shrinks to zero size at the horizon, the tip of the cigar.  One may reinterpret this geometry as a 2+1 dimensional confining gauge theory by Wick rotating one of the spatial coordinates back to be the time direction of this 2+1 dimensional gauge theory.  Then the metric in Eq. (\ref{AdSSch}) becomes
\begin{equation}
 u^2 (-h dt^2 + dx^2 + dy^2 + dz^2) +
    {du^2 \over h}  \, \rightarrow \,   u^2 ( h d\tau^2 + dx^2 + dy^2 - dt^2) + {du^2 \over h} .
\end{equation}
The same strategy can also be used to construct a gravity dual for a 3+1 dimensional confining gauge theory \cite{Witten:1998zw}. One simply reinterprets the gravity dual of maximally supersymmetric YM theory in 4+1 dimensions at finite temperature, which is given in terms of the near horizon geometry of near-extremal D4 branes in type IIA string theory, as a confining gauge theory in 3+1 dimensions. The gauge theory has a tower of KK-modes in addition to a theory of pure glue in 3+1 dimensions. In the weak coupling limit one describes pure glue; in the limit where the gravity dual is weakly curved, the KK-modes have masses of order $\Lambda_{\rm QCD}$. The resulting metric is ($i$, $j$ run over the 3 spatial directions, $x_4$ is the compact circle to get from 5 to 4 dimensions)
\begin{equation}
\nonumber
ds^2 = \left ( \frac{u}{R_{D4}} \right )^{3/2} \left [ - dt^2 + \delta_{ij} dx^i dx^j + f(u) dx_4^2
 \right ] + \left ( \frac{R_{D4}}{u} \right )^{3/2} \left [ \frac{du^2}{f(u)} + u^2 d \Omega_4^2 \right ],
 \end{equation}
where $f(u) = 1 - \left ( \frac{u_{\Lambda}}{u} \right )^3$ and $u_{\Lambda}$ is the minimum value of the $u$-coordinate, the tip of the cigar. There is also a scalar field turned on: the 5d dilaton has a profile $e^{\phi} = g_s \left ( \frac{u}{R_{D4}}\right)^{3/4}$. The $x_4$ direction is periodic with period
\begin{equation}
2 \pi R = \frac{4 \pi}{3} \left ( \frac{R^3_{D4} }{u_{\Lambda}} \right)^{1/2}.
\end{equation}
In the field theory, $R$ is the radius of the circle one compactified 4+1 SYM on, $g_5^2 = (2 \pi)^2 g_s l_s$ is the 5d YM coupling, and correspondingly $g_4^2 = g_5^2/(2 \pi R)$ is the 4d YM coupling.

To add fundamental flavors to this geometry, we can follow the approach described in Section \ref{flavor}. Two obvious candidates are probe D6-branes \cite{Kruczenski:2003uq} or probe D8-branes \cite{Sakai:2004cn} with four and six ND directions respectively.  The former preserves eight supersymmetries of the 5d super Yang-Mills. But supersymmetry in the flavor sector is broken by the compactification with anti-periodic boundary conditions for the fermions, just as it was in the glue sector. Chiral symmetry ensures massless fermions in 4d, but the scalar superpartners become massive via loop corrections. In the weak coupling limit one obtains real QCD with fermionic quarks only. In the strong coupling limit one has even more extra matter with masses of order the strong coupling scale: the KK-modes of the flavor fermions and their scalar partners. The advantage of using the D8-brane probe, with six ND directions, is that one introduces only fermions. A single D8 introduces a chiral fermion.  Anomaly freedom requires that the D8 is accompanied by an anti-D8. This system is stable as long as the D8's are separated in the direction of the compactified circle.

Both of the systems described in \cite{Kruczenski:2003uq,Sakai:2004cn} nicely exhibit chiral symmetry breaking of the axial $U(1)$ symmetry (whose anomaly is negligible at large $N$).  The low energy excitations of the flavor branes are the mesons \cite{Kruczenski:2003be}, and hence the DBI action plays the role of the chiral Lagrangian. The D8 model has the disadvantage that it is difficult to add a mass term for the fermions. As fermions of different chirality live at different places in the fifth dimension, the mass operator is non-local. To some extent the effects of such a non-local operator can be captured via the standard dictionary \cite{Aharony:2008an}. The advantage of the D8 system is that for more than one flavor one can observe chiral symmetry breaking of the full non-abelian chiral $SU(N_f) \times SU(N_f)$ to its diagonal subgroup in a very beautiful geometric manner: deep in the IR of the geometry the D8 and anti-D8 reconnect. Due to confinement, the D8's cannot end in the IR and have no option other than to connect.  Thus one sees in this model that confinement implies chiral symmetry breaking. The D8 system also has an additional parameter, the separation of the 4d flavors along the compactification circle.  This allows one to dial the physics in the meson sector independently of the glue sector. In particular, the typical masses in the meson sector are set by $L$, the asymptotic separation of the D8 and anti-D8, while the glueball masses are set by $R$, the compactification radius. For $1/L \gg 1/R$ the mesons are much heavier than the glueballs. For $1/L \gg 1/R$ one also has a very interesting phase structure where in the temperature window between $1/R$ and $1/L$ the theory is deconfined, but chiral symmetry is still broken. The D8 system, often referred to as holographic QCD (HQCD) or the Sakai-Sugimoto model, is currently the closest to QCD we have come with a theory that has a gravity dual.

\subsection{The warped deformed conifold or Klebanov-Strassler Background}
\label{ks}

Another popular supergravity solution dual to a confining gauge theory is the warped deformed conifold of Klebanov and Strassler \cite{Klebanov:2000hb}. The dual gauge theory is supersymmetric and exhibits (discrete) chiral symmetry breaking even before adding additional flavors via probe branes. The UV of this theory is very interesting: the dual field theory has a product gauge group out of which one factor is asymptotically free, while the other runs to strong coupling in the UV.  As this second gauge factor approaches its Landau pole, the gauge bosons turn out to be better described as composites of a different
gauge group. Using these dual variables the roles of asymptotically free and non-free gauge group factors change and one can run further up in energy until the other gauge coupling becomes strong and the same story repeats itself. Like in our other examples, one can take a limit in which this theory reduces to pure glue in the IR, but this involves string scale curvatures in the bulk. The limit in which supergravity is applicable has again extra matter of mass of order $\Lambda_{\rm QCD}$ whose net effect is to push up the flux tube tension.  Reviews have been written about the warped deformed conifold and its applications to field theory and cosmology: see for example
\cite{Herzog:2002ih,Benna:2008yg}.

\subsection{Confinement/Deconfinement transition}

One more universal feature of confining theories with a gravity dual which we haven't yet discussed in detail is their phase structure as a function of temperature.  Three-color QCD with 2 light flavors and a strange quark can have a first order transition, a second order transition, or a smooth crossover from hadronic degrees of freedom at low temperature to a gas of quarks and gluons at high temperatures depending on the precise values of the quark masses (with real world QCD exhibiting a crossover according to lattice simulations). In contrast to this, confining large $N$ glue theories are expected to always have a first order transition separating the hadronic phase from the quark-gluon phase. The reason for this is that the $N$ scaling of the free energy is very different in the two phases. The confined hadrons have order 1 degrees of freedom, whereas the deconfined glue has order $N^2$. The order $N^2$ part of the free energy is identically zero at low temperatures, so that the derivative of the free energy has to be discontinuous at the critical temperature $T_c$ where we go over from hadron gas to gluon plasma. The presence of fundamental matter does not alter this conclusion as long as the number of flavors $N_f$ is finite. In the deconfined phase, the quarks contribute a subleading order $N_f N$ contribution to the free energy. Only when $N_f \sim N$ can one get a smooth crossover as in the real world. In this case the $N_f^2$ mesons in the confined phase can smoothly connect with the $N^2$ gluons in the deconfined phase.

In the gravity dual, this phase transition is realized as a competition between two geometries \cite{Witten:1998zw}. One possibility for the gravity dual to a confining theory at finite temperature is to take any of the metrics we described above with the Euclidean time periodically identified.  Another option is to introduce a black hole horizon which completely hides the wall behind it: in Euclidean signature this means smoothly ending the geometry before the wall is encountered. To see which of those two geometries dominates the thermal ensemble one can compare their free energies. At low temperature the background with wall dominates (the hadronic phase), at large temperature the black hole (the deconfined phase). Out of the confining examples discussed above, an analytic solution describing the high temperature black hole solution is only available for Witten's black hole: the high temperature solution is the same as the low temperature one, just with the roles of spatial and time circle exchanged.  Numerical black hole solutions have been found for the warped deformed conifold \cite{Aharony:2007vg} and the ${\cal N}=2^*$ solution \cite{Buchel:2007vy}. One pleasant feature of this transition is that the high temperature phase is insensitive to the details of the confinement mechanism. Many features of the high temperature phase are shared between all the confining theories with gravity dual and, in fact, also the original ${\cal N}=4$. Most studies of finite temperature QCD using the gauge-string duality use the ${\cal N}=4$ theory itself, as this is expected to be as good a model of the quark-gluon plasma as any of the confining examples.

\section{Shear and bulk viscosity}
\label{VISCOSITY}
\subsection{Entropy and shear viscosity}

Special interest attaches to the ratio $\eta/s = 1/4\pi$ of shear viscosity to entropy density predicted by the gauge-string duality \cite{Policastro:2001yc}.  On the theoretical side, this calculation is appealing because of its universality \cite{Buchel:2003tz}, and because of the conjecture \cite{Kovtun:2004de} that $\eta/s \geq 1/4\pi$ for any physically reasonable medium.  On the experimental side, it is interesting because elliptic flow in heavy-ion collisions seems to be best understood in terms of nearly ideal fluid dynamics, with $0 < \eta/s \mathrel{\mathstrut\smash{\ooalign{\raise2.5pt\hbox{$<$}\cr\lower2.5pt\hbox{$\sim$}}}} 0.2$, a range clearly consistent with the value $1/4\pi$ from the gauge-string duality in the strong coupling limit.  A recent review centered on the $\eta/s$ calculation has appeared in Ann.\ Rev.\ Nucl.\ Part.\ Sci.\ \cite{Son:2007vk}, to which we refer the reader both for a more thorough exposition and for a more extensive set of references.

To understand the result $\eta/s = 1/4\pi$ in a pedestrian fashion, let's start by considering the computation of the entropy of near-extremal D3-branes.  More precisely, let's start with the $AdS_5$-Schwarzschild metric, (\ref{AdSSch}), times a five-sphere $S^5$ of radius $L$.  The horizon is at $u=u_h$.  The $dt^2$ part of the metric vanishes at this value of $u$, and the $du^2$ part becomes singular.  In a more proper treatment, the $t$-$u$ plane would be parametrized instead by Kruskal coordinates, and the metric would be seen to be perfectly smooth at the horizon.  What matters for us, however, is that the $dx_1^2+dx_2^2+dx_3^2$ part of the metric is smooth at $u=u_h$, and that the $S^5$ part of the metric is entirely independent of $u$.  The metric on the horizon is just
 \begin{equation}
  ds_{\rm horizon}^2 = L^2 u_h^2 (dx_1^2 + dx_2^2 + dx_3^2) +
    L^2 d\Omega_5^2 \,.
    \label{HorizonMetric}
 \end{equation}
What this means is that the area $A_h$ of the horizon is just $V_3 L^3 u_h^3 \Vol S^5$, where $\Vol S^5 = \pi^3 L^5$ is the volume of the five-sphere and $V_3$ is the coordinate three-volume in the $x_1$, $x_2$, and $x_3$ directions.  Using the relation $S = A_h / 4G_{10}$ between the entropy $S$ and horizon area $A_h$, we arrive at
 \begin{equation}
  s \equiv {S \over V_3} = {\pi^3 L^8 u_h^3 \over 4 G_{10}} \,.
    \label{GotS}
 \end{equation}
To obtain the familiar result $s = {3 \over 4} s_{\rm free}$ of \cite{Gubser:1996de}, one needs the relation $T = u_h/\pi$ and also the relation $L^8 / G_{10} = 2N^2/\pi^4$, which is specific to ${\cal N}=4$ super-Yang-Mills and $AdS_5 \times S^5$ and originates in the formula for the tension of D3-branes.

Next we need to understand the gravity calculation that determines $\eta$.  Suppose we stand far away from a near-extremal stack of D3-branes, in asymptotically flat space, and throw gravitons at it.  When the energy $\omega$ of the gravitons is much less than $1/L$, where $L$ is the radius of curvature characteristic of the D3-branes, the absorption will be dominated by the $s$-wave component of the incoming flux of gravitons.  A famous result \cite{Das:1996we} is that the absorption cross-section equals the horizon area in the extreme low-energy limit:
 \begin{equation}
  \sigma_{\rm absorb} = A_h \,. \label{SigmaResult}
 \end{equation}
The result (\ref{SigmaResult}) is intuitive: surely the $s$-wave cross-section would be at least proportional to the horizon area.  But it is not inevitable that the constant of proportionality should be one.  This factor comes from first showing that the relevant linearized Einstein equations reduce to the massless Klein-Gordon equation in the black hole background, and then analyzing the low-frequency solutions of the Klein-Gordon solution with purely infalling boundary conditions at the horizon.

To connect (\ref{SigmaResult}) with shear viscosity, one needs three ingredients.  First, we have Kubo's formula:
 \begin{equation}
  \eta = -\lim_{\omega \to 0} {1 \over \omega} \Im G_R(\omega) \,,
    \label{Kubo}
 \end{equation}
where
 \begin{equation}
  G_R(\omega) = -i \int d^3 x \int_0^\infty dt \,
    \langle [ T_{12}(t,\vec{x}), T_{12}(0,0) ] \rangle
    \label{GRdef}
 \end{equation}
is the retarded Green's function of the $12$ component of the stress tensor.  (Any traceless spatial part of $T_{mn}$ would do as well here.)  Next, we have the spectral representation
 \begin{eqnarray}
  \Im G_R(\omega) &=& \sum_{i,f}
    {e^{-\beta E_i} \over Z}
     \left| \langle f | T_{12}(0,0) | i \rangle
      \right|^2 (2\pi)^3 \delta^3(\vec{p}_f - \vec{p}_i)  \nonumber \\
    &&\qquad{} \times
      \pi \left[ \delta(E_f-E_i+\omega) - \delta(E_f-E_i-\omega)
        \right] \,.
 \end{eqnarray}
Finally, we have an application of Fermi's Golden Rule to compute the net rate at which gravitons are absorbed by the branes from a plane wave of gravitons polarized so that only $h_{12}$ is non-vanishing.  The interaction of the gravitons with the branes is
 \begin{equation}
  S_{\rm int} = \int d^4 x \, h_{12} T^{12} \,.  \label{Sint}
 \end{equation}
Far from the D3-branes, in asymptotically flat space, let's take $h_{12} = e^{-i\omega (t - x_4)}$, where $x_4$ is one of the directions normal to the brane.  Then the flux of gravitons is
 \begin{equation}
  {\cal F} = {\omega \over 8\pi G_{10}} \,.  \label{Fdef}
 \end{equation}
The $\omega$ arises from the usual structure of fluxes for bosons: ${\cal F} = {1 \over 2i} \phi^* \overleftrightarrow\partial \phi$ for a canonically normalized complex boson $\phi$.  But $h_{12}$ isn't canonically normalized, due to the factor $1/16\pi G_{10}$ in front of the Einstein action in ten dimensions.  The $8\pi G_{10}$ in the denominator of (\ref{Fdef}) compensates for this, as well as the fact that we must take the real part of $h_{12}$ at the end of the day.  Fermi's Golden Rule says that the net absorption rate of gravitons is
 \begin{eqnarray}
  \Gamma &=& V_3 \sum_{i,f} {e^{-\beta E_i} \over Z}
    \left| \langle f | T_{12}(0,0) | i \rangle \right|^2
    (2\pi)^3 \delta^3(\vec{p}_f - \vec{p}_i)  \nonumber \\
     &&\qquad{} \times 2\pi \left[ \delta(E_f - E_i - \omega) -
      \delta(E_f - E_i + \omega) \right]  \nonumber \\
    &=& -2 V_3 \Im G_R(\omega) \,.  \label{Golden}
 \end{eqnarray}
On the other hand, the defining relation for $\sigma_{\rm absorb}$ is $\Gamma = \sigma_{\rm absorb} {\cal F}$.  Combining this relation with (\ref{Kubo}-\ref{Golden}), one arrives at
 \begin{equation}
  \eta = \lim_{\omega \to 0} {\sigma_{\rm absorb} / V_3 \over
    16\pi G_{10}} = {A_h / V_3 \over 16\pi G_{10}} = {s \over 4\pi}
     \,.  \label{FinalVisc}
 \end{equation}
In the second equality of (\ref{FinalVisc}), we used the result (\ref{SigmaResult}).  In the third equality, we used the relation $S = A_h / 4G_{10}$.  Note that we did not require the explicit computation (\ref{GotS}).

The universality of the result $\eta/s=1/4\pi$ may be seen as approximately parallel to the universality of the result $\sigma_{\rm absorb} = A_h$.  More explicit checks of universality have been performed \cite{Buchel:2003tz,Buchel:2004qq,Iqbal:2008by}.  Corrections from $\alpha'$ effects have been calculated: first, apparently incorrectly, in \cite{Buchel:2004di}, and later, apparently correctly, in \cite{Buchel:2008ac}. Loop effects, corresponding to $1/N$ corrections in the dual gauge theory, have been explored as well in \cite{Myers:2008yi}, and in \cite{Kats:2007mq} it was shown that the bound $\eta/s \geq 1/4\pi$ is violated---perhaps only slightly---by a string theory construction involving D3-branes at an orientifold singularity. The latter calculation has been checked and confirmed in \cite{Buchel:2008vz}.

\subsection{Bulk viscosity}

Bulk viscosity, denoted $\zeta$, quantifies energy loss due to expansion.  It can be described in terms of the constitutive relations for relativistic hydrodynamics:
 \begin{eqnarray}
  T^{mn} &=& (\epsilon+p) u^m u^n + p g^{mn} -
      \nonumber \\ &&\quad{}
    P^{ma} P^{nb} \left[
     \eta \left( \partial_a u_b + \partial_b u_a
       - {2 \over 3} g_{ab} \partial_c u^c
       \right) +
     \zeta g_{ab} \partial_c u^c \right] \,. \label{Constitutive}
 \end{eqnarray}
Here $u^m = (\gamma, \gamma \vec{v})$ is the local four-velocity, and $P^{mn} = g^{mn} + u^m u^n$ is the projector onto the spatial directions of the local rest frame.  Indices $m,n$ run from $0$ to $3$, while indices $a,b$ run from $1$ to $3$.  Conformal invariance is summarized by the identity $T^m{}_m = 0$, and inspection of (\ref{Constitutive}) shows that this demands the familiar relation $p = \epsilon/3$ and also $\zeta=0$.  But conformal symmetry may be broken in the gauge-string duality: indeed, there is a large literature studying renormalization group flows, and all the gravity duals of confining gauge theories are at some level examples of renormalization group flows.  A number of authors have considered bulk viscosity computations in non-conformal gauge-string backgrounds \cite{Parnachev:2005hh,Mas:2007ng,Buchel:2008uu}.

One of the main guidelines to the output of such computations is the bound \cite{Buchel:2007mf}
 \begin{equation}
  {\zeta \over \eta} \gsim 2 \left( {1 \over 3} - c_s^2 \right) \,.
 \end{equation}
Computations of $\zeta$ in theories where the equation of state is close to that of QCD have appeared in \cite{Gubser:2008yx,Gubser:2008sz}; see also the closely related work \cite{Gursoy:2008bu}, where the equation of state is closer to that of pure Yang-Mills theory.  The outcome of these calculations is that $\zeta/s$ does reach a maximum in the middle of the deconfinement crossover, but its maximum value is approximately bounded by $1/4\pi$, which means that bulk viscosity is not a dominant source of entropy production in heavy ion collisions.  There is some tension between this result and the results of \cite{Kharzeev:2007wb}, whose sum-rule calculations suggest that bulk viscosity becomes the dominant correction to inviscid hydrodynamics near freeze-out.  Lattice calculations on pure Yang-Mills \cite{Meyer:2007dy} find values of $\zeta/s$ as high as $0.7$ very close to $T_c$.  It may be perilous, however, to make direct comparisons between pure Yang-Mills theory and QCD for quantities evaluated right near $T_c$, given that one theory goes through a first order phase transition while the other goes through a cross-over.

\section{Expanding Plasmas}
\label{EXPANDING}

As with perturbative methods, it is much simpler to study static finite-temperature configurations in the gauge-string duality than to study expanding plasmas.  But the expansion is obviously a crucial feature of experimental conditions.  A notable attempt to go beyond static configurations is the work \cite{Janik:2005zt}, in which an approximate holographic dual of Bjorken flow was found.  This flow has boost invariance in the $x_3$ direction and translation invariance in the $x_1$ and $x_2$ directions.  It describes a plasma which is infinite in all three spatial directions and expands in the $x_3$ direction, with a local velocity $v_3 \propto x_3$ at any fixed time $t>0$.  The late-time behavior found in \cite{Janik:2005zt} is precisely what follows from inviscid hydrodynamics; however, a more refined analysis \cite{Janik:2006gp} shows that shear viscosity corrections enter with a magnitude as expected from the relation $\eta/s = 1/4\pi$.  This is reassuring, because the original derivation \cite{Policastro:2001yc} was for perturbations around a static configuration, contrasting with the dynamical setting of Bjorken flow.  Subsequent work, including \cite{Bhattacharyya:2008jc}, confirms earlier suggestions, for example in \cite{Policastro:2002se}, that there is a general correspondence between non-static black hole horizons and fluid dynamics in a dual gauge theory.  The general flavor of this connection is to develop the dynamics of the black hole horizon in a long-wavelength expansion.  The earliest terms in such an expansion reveal thermodynamic properties and inviscid fluid dynamical behavior; higher order corrections allow the extraction not only of the shear viscosity, but also of dynamical transport coefficients that characterize the response to higher derivative deformations of uniform flow profiles. These transport coefficients can also be extracted from a sufficiently precise analysis of linearized perturbations of the $AdS_5$-Schwarzschild solution \cite{Baier:2007ix}.

Another line of thought \cite{Friess:2006kw} is to describe a finite-sized, expanding plasma in terms of a static black hole in global anti-de Sitter space.  This proposal follows earlier work \cite{Horowitz:1999gf}, and it leads to exact solutions for the expectation $\langle T_{mn} \rangle$ of the dual stress tensor.  A disadvantage is that the exact solutions have perfect radial symmetry.  They describe a spherical shell of conformal plasma collapsing to a minimum radius and then re-expanding.  There is no shear and no entropy production.  However, one can perturb the exact solutions to study both hydrodynamical deformations and the rate of local equilibration to a hydrodynamical description.  On the dual black hole side the perturbations are quasi-normal modes (QNM's).  As was already understood in \cite{Kovtun:2005ev}, a small subset of the QNM's describe perturbations in linearized hydrodynamics, while the others describe non-hydrodynamical behaviors.  The rate of decay of the non-hydrodynamical modes is quite fast: their $e$-folding time is at most $\tau_{e-\rm fold} \sim 1/8.6 T_{\rm peak}$, where $T_{\rm peak}$ is the maximum temperature of the equilibrated plasma.  In \cite{Friess:2006kw} it was suggested that, given this $e$-folding time, it would take a total time $\tau \approx 0.3\,{\rm fm}/c$ for the strongly anisotropic momentum-space distribution of the initial state to equilibrate to a locally thermal state under conditions similar to a gold-gold collision at $\sqrt{s_{NN}} \approx 200\,{\rm GeV}$.  A similar point was made in \cite{Janik:2006gp} in the context of the holographic dual to Bjorken flow.  Related estimates have appeared in \cite{Hod:2006jw,Bak:2007qw}.  The strong-coupling estimates are significantly shorter than straightforward estimates from pQCD, which---according to \cite{Arnold:2004ti}---are typically in the range $2.5\,{\rm fm}/c$ or higher.

Yet another related line of thought is to relate the formation of a black hole horizon in a high-energy collision in $AdS_5$ to the production of entropy in a heavy-ion collision.  Contributions in this area include the works \cite{Nastase:2005rp}, which hinges on a proposed relation between gravitational shock waves in a cut-off version of $AdS_5$ and a coherent pion field; \cite{Shuryak:2005ia}, which proposes a translation of the whole history of a heavy-ion collision into gravitational language; \cite{Gubser:2008pc}, which provides a quantitative estimate of the total multiplicity in gold-gold collisions; and \cite{Grumiller:2008va}, which includes a study of the metric shortly after the collision as well as an estimate of the thermalization time which, like the ones discussed in the previous paragraph, is short compared to perturbative estimates.  Recently, the formation of a black hole in $AdS_5$ in response to a time-dependent metric on the boundary was studied numerically \cite{Chesler:2008hg}.  The dual gauge theory equilibrates to an effectively thermal state in a time which is short compared to perturbative estimates, indicating at least qualitative agreement with the analysis of QNM's.

\section{Hard Probes and Heavy Mesons}
\label{PROBES}
\subsection{Hard Probes}

In QCD, the coupling runs as a function of scale. In relativistic heavy ion collisions, different sub-processes happen at different scales and hence at different couplings.  Most of the collisions apparently lead to end-products which quickly thermalize and produce the expanding fireball of strongly-coupled quark-gluon plasma. But some of the partons collide basically head on, leading to standard perturbative QCD scattering, which produces very energetic partons. Even though these energetic partons originate in the same nucleus-nucleus collision, they are better thought of as external probes traveling through the plasma.  In fact these ``hard probes'' serve as an excellent tool to study the plasma.  The way they interact with the plasma and the rate at which they deposit energy and momentum reveal detailed properties of the plasma.

In order to get a theoretical handle on the properties of hard probes, one first wants to understand the idealized scenario of a probe traveling through an infinitely extended, spatially uniform plasma. The resulting transport properties can than be included in hydrodynamic codes modeling the real fireball in a heavy ion collision. At weak coupling, the energy loss by heavy and light quarks or gluons is well understood in QCD, as we partly review below.  In the regime of the strongly coupled quark-gluon plasma, we are lacking any reliable theoretical tools, so getting some guidance from gauge/gravity models to the qualitative features of such probes is clearly desirable.

As we discussed, in the simple example of ${\cal N}=4$ SYM the coupling does not run as a function of scale, and so one does not automatically produce hard partons in a high-energy collision. This point has been made most explicitly in the paper \cite{Hofman:2008ar}. There it was shown that at zero temperature a single off-shell photon produces a completely round energy distribution. This is in stark contrast to QCD where the same process (which experimentally is realized in DIS) for an off-shell photon with energy far above the strong coupling scale produces a jet-like antenna pattern, that is energy sharply localized along the trajectories one would
want to associate to a ``back-to-back'' quark/anti-quark pair produced in a perturbative QCD process. In strongly coupled ${\cal N}=4$ SYM it simply produces many very soft gluons.

In order to circumvent this issue in the study of hard probes using the gauge/ gravity correspondence, two main routes have been pursued. One is to augment ${\cal N}=4$ SYM with additional very heavy flavors using the technique describe in section \ref{flavor}. As long as the mass is sufficiently large compared to the temperature,\footnote{In QCD the requirement would simply be that the quark mass be much larger than the temperature, but in theories with a gravity dual the meson mass is parametrically smaller than the quark mass due to the large binding energy. In this case the relevant criterion for whether a quark is heavy is whether the meson masses are much larger than the temperature. In the ${\cal N}=4$ case this for example implies that the quark mass is much larger than $\sqrt{\lambda} T$, not just T.} the heavy quark will automatically serve as a hard probe of the plasma.  Its energy loss is well described by the ``trailing string'' of refs \cite{Herzog:2006gh,Gubser:2006bz} as we will describe below: see also the closely related calculation of \cite{CasalderreySolana:2006rq}.  An earlier approach to energy loss from hard probes \cite{Sin:2004yx}, loosely related to the trailing string, is based on the gauge-string treatment \cite{Mikhailov:2003er} of radiation from an accelerating charge.  For the trailing string, not just the rates, but the full energy and momentum profile of the wake left behind by the probe are well understood.

Another route in the string theoretic study of hard probes is to relax the requirement that one must describe an entire scattering process in terms of the gauge-string duality. One appeals to the weakly coupled regime of QCD for the initial production of the hard probe and then only studies the ensuing time evolution of the hard probe in the strongly coupled ${\cal N}=4$ plasma. One example in this spirit is the calculation of the jet quenching parameter of \cite{Liu:2006ug}, where it is first argued that in QCD the energy loss of a highly relativistic parton is given in terms of a light-like Wilson loop and then only the latter is calculated using the gauge gravity correspondence. Another example is the falling strings of refs \cite{Gubser:2008as,Chesler:2008wd,Chesler:2008uy} where one sets up in the ${\cal N}=4$ SYM an initial state that corresponds to a sharply localized initial distribution of energy and momentum, as from an energetic light quark or gluon.
The subsequent evolution of such a distribution in the ${\cal N}=4$ plasma can be studied completely within the framework of the gauge-string duality. The production of such an initial state is quite unnatural in strongly coupled ${\cal N}=4$ SYM. But such states are of course typical in weak coupling QCD scattering.  Once produced, their subsequent evolution should be similar in strongly coupled QCD and strongly coupled ${\cal N}=4$.

Let us describe some of the constructions in more detail. As we argued above, the simplest question to address is energy loss by a heavy quark. In this case the problem can be entirely phrased within the ${\cal N}=4$ theory. Heavy flavors can easily be incorporated using the D7 probes of section \ref{flavor}.  On the supergravity side, one is looking for a string profile that has a single endpoint on the brane moving along at a constant velocity $v$. As the quark loses energy and momentum at a constant rate due to the drag force experienced in the plasma, one needs to turn on a background electric field on the brane in order to ensure that the quark can maintain constant velocity; or one can work in a formal limit where the quark is infinitely heavy, and so maintains constant $v$ despite a finite rate of energy loss. Asking for a stationary embedding of the string worldsheet of the form $x(t,z) = v t + f(z)$, where $x$ is the direction of motion the quark, $z$ is the radial $AdS_5$ coordinate and $t$ is time, the minimal area requirement gives a single ODE for $f(z)$. Furthermore, since the metric is independent of $x$, the area of the worldsheet only depends on $f'$ and not on $f$, and so the equations of motion following from the minimal-area action for the string can simply be integrated. The generic solution describes a string that turns around and has a second endpoint on the brane/boundary: a quark/antiquark pair. There is a unique choice for the integration constant in $f'$ that leads to a string with only one endpoint at the brane. In the IR this ``trailing'' or ``dragging'' string logarithmically approaches the horizon.\footnote{As explained in \cite{CasalderreySolana:2007qw}, it crosses the horizon when described using Kruskal coordinates.} Energy and momentum flow down at a constant rate down the string.  This
corresponds to a redistribution of energy along the string, from parts of the string close to the endpoint (which are best thought of as part of the quasi-particle consisting of the quark and the surrounding gluon cloud) to parts of the string close to the horizon (which are best thought of as part of the plasma). In the case of the trailing string, the induced metric on the worldsheet is a 2d black hole \cite{Gubser:2006nz,CasalderreySolana:2007qw}, and so a natural choice would be to consider the worldsheet horizon as the boundary between the two regions, but in general there is some ambiguity in how to separate the string into quasi-particle and plasma. This is of course in complete analogy with field theory, where the definition of a quasi-particle as a localized lump of energy and charge suffers from a similar ambiguity.

A series of papers \cite{Friess:2006aw,Friess:2006fk,Yarom:2007ap,Gubser:2007nd,Yarom:2007ni,Gubser:2007xz,Chesler:2007an,Gubser:2007ga,Chesler:2007sv} explains how to calculate the spatial distribution of energy loss starting from the trailing string.  In the infrared, there is a good match to hydrodynamics with $\eta/s = 1/4\pi$.  The sonic boom and the diffusion wake are of comparable strength in the hydrodynamic regime, in line with earlier hydrodynamic calculations, in particular ``Scenario 1'' of \cite{CasalderreySolana:2004qm}.  It has been suggested \cite{Stoecker:2004qu,CasalderreySolana:2004qm} that a sonic boom is responsible for the experimental observations summarized as ``jet-splitting:'' namely, a broad peak in di-hadron histograms consistent with high-angle emission from an energetic parton traversing the quark-gluon plasma \cite{Adare:2008cq,Ulery:2005cc}.  It appears, however, to be challenging to get a strong enough sonic boom to make this idea work \cite{Chaudhuri:2005vc}.  It is therefore interesting to note that analyses \cite{Noronha:2008un,Betz:2008wy} combining the flow patterns calculated from the trailing string with Cooper-Frye hadronization \cite{Cooper:1974mv} show a significant jet-splitting effect from region close, but not too close, to the quark.  Comparison of the results of \cite{Noronha:2008un,Betz:2008wy} with data is complicated by the facts that expansion of the plasma was neglected, and that the trailing string describes heavy quarks, which presumably make a very small contribution to untagged di-hadron histograms.

The energy loss experienced by a light quark is a little more complicated. In QCD energy loss by an ultra-relativistic parton can be determined in the BDMPS framework of \cite{Baier:1996sk,Baier:2001yt}, which describes radiative energy loss. The important parameter that sets the loss rate is the jet quenching parameter $\hat{q} = \langle p_{\perp}^2 \rangle/x$, the average transverse momentum acquired by a parton after it has traveled a distance $x$, measured in the rest frame of the plasma. This calculation is based on perturbative QCD, and therefore this framework is certainly not applicable when discussing the energy loss of relativistic partons in strongly coupled ${\cal N}=4$ SYM, whose coupling remains
constant even at high energies. It was however argued in \cite{Liu:2006ug,Liu:2006he} that $\hat{q}$ can be rephrased as an expectation value of a partially lightlike Wilson loop. This Wilson loop then can be determined in AdS/CFT via a minimal area calculation. Some doubts remain whether the saddle point identified in \cite{Liu:2006ug} is the dominant contribution to the bulk path-integral, see for example \cite{Argyres:2006vs,Argyres:2006yz}.
For heavy quarks, the jet quenching parameter---if defined as $\hat{q} = \langle p_\perp^2 \rangle / x$---can be directly determined in AdS/CFT \cite{Gubser:2006nz,CasalderreySolana:2007qw}.  In fact, both transverse and longitudinal momentum dispersion can be determined in terms of small fluctuations around the trailing string described above.  For a heavy quark with boost factor $\gamma$, the transverse momentum broadening grows as $\gamma^{1/2}$, while the longitudinal momentum broadening grows with $\gamma^{5/2}$.  So formally, in the limit of an ultra-relativistic particle the jet-quenching parameter diverges.  This is not in direct contradiction to the result of \cite{Liu:2006ug}, as the analysis of the trailing string based on classical worldsheets is expected to receive quantum corrections for $\gamma$'s that are parametrically large compared to $\lambda$, whereas the calculation of $\hat{q}$ based on lightlike Wilson lines is best thought of being done at infinite $\gamma$ for large but finite $\lambda$.  Last but not least, one can directly study the energy loss of a light quark or gluon in the strongly coupled ${\cal N}=4$ directly. In this case one is studying a falling string in the
$AdS_5$-Schwarzschild geometry. Analytic estimates of such falling strings from ref \cite{Gubser:2008as} revealed that the stopping distance of a light quark is expected to scale with the energy of the quark as $E^{1/3}$.  This scaling has also been seen analytically and numerically in \cite{Hatta:2008tx,Chesler:2008wd,Chesler:2008uy}.  \cite{Chesler:2008uy} also revealed that most of the energy of the light quark is released in a burst of energy at the very end of the trajectory, a phenomenon that is well known as a ``Bragg peak" in other contexts.

\subsection{Heavy Flavor and Mesons}

So far we have been discussing the properties of colored hard probes traversing the medium. For certain physical phenomena, such as J/$\psi$ suppression, it is also important to understand the properties of color neutral mesons moving through the plasma. While the high temperature phase of QCD is a theory of free quarks and gluons, there is some evidence from the lattice (for a review see \cite{Satz:2005hx}) that bound-states involving heavy flavors such as charm survive to temperatures above the crossover at least as peaks in the spectral density. In a generic finite temperature theory, there can never be sharp bound-states at finite temperature; thermal fluctuations always lead to a finite width. However, in the limit of a large number of colors that underlies the classical supergravity limit of gauge-string constructions, heavy flavor mesons can survive for some range of temperatures as true bound states with a width of order $N_c^{-1/2}$ that formally goes to zero. So the gauge-string duality is a perfect testing ground for mapping out properties of these heavy flavor mesons at finite temperature.

In the study of flavors via probes branes, it has been found that a generic flavored gauge theory with gravitational dual exhibits a meson melting phase transition as a function of temperature. For temperatures below a critical value set by the mass of the lightest heavy flavor meson, the mesons are stable. As the temperature is increased they fall apart, and even heavy flavor only exists as free quarks. This meson melting happens in a sharp first order phase transition with a kink in the free energy. This phenomenon has been first observed for the D3/D7 system in \cite{Babington:2003vm,Kirsch:2004km} and then has been shown to be a generic outcome of a probe brane analysis in \cite{Mateos:2006nu}. The fact that the transition is first order is an artifact both of the limit of a large number of colors and large 't Hooft coupling and it has been argued that both finite $N_c$ \cite{Dusling:2008tg} and finite coupling effects \cite{Faulkner:2008qk} will smooth out the transition.

Besides the existence of this meson-melting phase transition, another interesting generic feature of mesons from probe branes is the existence of a speed limit. This can be seen either from the analysis of the modes living on the brane \cite{Mateos:2007vn,Ejaz:2007hg}, or from the analysis of moving quark/anti-quark pairs as being described by semi-classical strings \cite{Peeters:2006iu,Herzog:2006gh,Liu:2006nn,Chernicoff:2006hi}. An important concept that allows one to see this effect in the dual bulk geometry is the local speed of light.  An object moving in the spatial direction along a slice of constant radial coordinate in the black hole background given by metric (\ref{AdSSch}) moves at the speed of light if its velocity is $h(u)^{1/2}$.  In other words, any object moving at a constant speed $v$ along the slice is actually superluminal if it reaches down to an effective horizon at
$v^2=h(u_h^{\rm eff})$.  Using $h = 1 - u_h^4/u^4$, one obtains $u_h^{\rm eff} = u_h/(1-v^2)^{1/4}$. For a flavor brane corresponding to heavy flavors with mass $m$ reaching down to a finite $u_m \sim \sqrt{\lambda} m$, mesons corresponding to a small fluctuation propagating along the brane with a speed $v$ experience an effective speed limit that requires $u_m > u_h^{\rm eff}$.\footnote{A phenomenological application has been proposed in \cite{CasalderreySolana:2008ne}: a flavorless meson, like the J/$\psi$, may produce an observable peak in the photon spectrum if its limiting in-medium velocity is subluminal.}  A similar speed limit is in effect for large spin mesons or a quark-antiquark dipole that is held up by an external field. In that case, the meson or dipole is described by a semi-classical string in the bulk that reaches down to some minimum turning point $u_0$ which is set by the separation between quark and antiquark. Again, the speed limit requires $u_0 > u_h^{\rm eff}$. While the details depend on the setup, the fact that the speed limit compares the typical energy scale of the excitation not to $T$ but to $T/(1-v^2)^{1/4}$ is universal.  Similar scaling can be observed in the fluctuation analyses of \cite{Gubser:2006nz,CasalderreySolana:2007qw}.

\section{Concluding remarks}
\label{CONCLUDE}

The dream of describing strong interactions in terms of string theory is an old one---older than QCD itself.  The gauge-string duality, as the modern incarnation of that dream, still falls short in important respects: asymptotic freedom is poorly understood, and gauge-string constructions typically lead to a parametric mismatch between the mass gap and the flux tube tension.  These defects are due to the almost universal reliance on the classical supergravity approximation, corresponding to the large $N$, large $g_{YM}^2 N$ limit of the dual gauge theory.  And yet, as we have reviewed, constructions using classical gravity and probe branes provide an excellent venue for calculations ranging from meson spectra to transport coefficients.  Finite temperature is elegantly included in these calculations in terms of a black hole horizon in the fifth dimension.  Making the comparisons between such calculations and QCD systematic is a persistent difficulty.  Despite this difficulty, gauge-string treatments of confinement and of finite-temperature non-abelian plasmas are quite valuable because they complement insights from more standard quantum field theoretic treatments.  For example, the descriptions of confinement are elegant and geometrical; the connection to hydrodynamics is relatively simple; and the interplay between hard probes and a thermal medium is rich and explicit.

Attempts to forecast the future of a field of theoretical physics usually are doomed to failure.  However, we look forward to future developments in the connections between AdS/CFT and strong nuclear interactions in several directions, including: studies of thermalization in terms of dual black holes; some better unification of the descriptions of energy loss; progress, if only incremental, toward finite 't~Hooft coupling; and better integration of the ``top-down'' constructions of confining duals with ``bottom-up'' models of holographic QCD.

\clearpage
\def\href#1#2{{\tt #2}}
\bibliographystyle{ssg}
\bibliography{pedestrian}

\end{document}